\documentclass[%
 reprint,
superscriptaddress,
 amsmath,amssymb,
 aps,
]{revtex4-2}

\usepackage{graphicx}
\usepackage{dcolumn}
\usepackage{bm}

\begin{document}

\preprint{APS/123-QED}

\title{Practical quantum imaging with undetected photons}

\author{Emma Pearce} 
\email{e.pearce18@imperial.ac.uk}
\affiliation{Blackett Laboratory, Department of Physics, Imperial College London, SW7 2AZ, United Kingdom}

\author{Nathan R. Gemmell}
\affiliation{Blackett Laboratory, Department of Physics, Imperial College London, SW7 2AZ, United Kingdom}

\author{Jefferson Fl\'{o}rez}
\affiliation{Blackett Laboratory, Department of Physics, Imperial College London, SW7 2AZ, United Kingdom}

\author{Jiaye Ding}
\affiliation{Blackett Laboratory, Department of Physics, Imperial College London, SW7 2AZ, United Kingdom}

\author{Rupert F. Oulton}
\affiliation{Blackett Laboratory, Department of Physics, Imperial College London, SW7 2AZ, United Kingdom}
 
\author{Alex S. Clark}
\affiliation{Blackett Laboratory, Department of Physics, Imperial College London, SW7 2AZ, United Kingdom}
\affiliation{Quantum Engineering Technology Labs, H. H. Wills Physics Laboratory and Department of Electrical and Electronic Engineering, University of Bristol, BS8 1FD, United Kingdom}

\author{Chris C. Phillips}
\affiliation{Blackett Laboratory, Department of Physics, Imperial College London, SW7 2AZ, United Kingdom}

\date{July 12, 2023}

\begin{abstract}
Infrared (IR) imaging is invaluable across many scientific disciplines, from material analysis to diagnostic medicine. However, applications are often limited by detector cost, resolution and sensitivity, noise caused by the thermal IR background, and the cost, portability and tunability of infrared sources. Here, we describe a compact, portable, and low-cost system that is able to image objects at IR wavelengths without an IR source or IR detector. This imaging with undetected photons (IUP) approach uses quantum interference and correlations between entangled photon pairs to transfer image information from the IR to the visible, where it can be detected with a standard silicon camera. We also demonstrate a rapid analysis approach to acquire both phase and transmission image information. These developments provide an important step towards making IUP a commercially viable technique.

\end{abstract}

\maketitle

\section{Introduction}

The infrared (IR) spectral region provides a wealth of information. In the near-IR and shortwave-IR (SWIR), higher harmonics of the vibrational modes of molecules and combinations of them can be probed. In the mid-IR, fundamental vibrational absorption bands occur, which provide both greater molecular specificity \cite{NIRreview,NIRbook} and the ability to perform quantitative chemical analysis. Sensing applications include studies of molecular structure, agriculture and food quality control, pharmaceutical monitoring, and biological imaging \cite{MolecularMixtures,NIRAgriculture,Brewing,Pharma,FishEggs,OralCancer,IRcancerreview}.
	
 However, the IR is technologically poor in comparison with the visible, particularly at longer MIR wavelengths. IR cameras have much lower pixel counts than their visible silicon counterparts and, when operated at room temperature, are orders of magnitude noisier. Even expensive cryogenically-cooled IR detectors  \cite{IRdetectors} are susceptible to noise arising from the ever-present 300\,K black-body radiation background (the so-called BLIP limit). 
 
 One approach to avoiding IR cameras is to image the sample with IR photons which then undergo frequency up-conversion to visible wavelengths before reaching the camera. However, this requires an IR source of photons and relies on high-power laser sources and/or cavities to combat the low conversion efficiencies \cite{Upcon:15,Upcon_video}. This also leads to the sample being exposed to far more photons than are detected, which can be detrimental to photosensitive samples \cite{PhotosensitiveReview,NathanBalancing}. So-called `ghost imaging' uses non-degenerate correlated photon pairs. The IR photon probes the sample  before being detected with a single-channel IR detector whilst its visible partner is logged by a camera \cite{Aspden:15,IR_GI}. This circumvents some of the limitations of IR cameras, but it still suffers from the poor IR detector sensitivity, and from the influence of thermal IR background. 

In contrast, the IUP approach circumvents both the requirement to have a direct source of IR photons and the ability to detect them \cite{ZWM1991,Lemos2014}. This works by generating photon pairs via spontaneous parametric down-conversion (SPDC) in a nonlinear crystal, where each pair consists of one visible photon (signal) and one infrared photon (idler). By passing the pump through this crystal twice, it is possible to generate a pair in the first and/or second pass of the pump. If the signal photons from these two passes are precisely overlapped, and the idler photons are similarly overlapped, it is impossible to determine which pass generated the photon pair. This lack of `which-way' (or indeed, `which-pass') information means that optical interference is seen in the count rate of the signal photons. Blocking one of the idler beams with an object effectively restores this `which-way' information, destroying the interference. Note that coincident detection is not required, as merely the possibility to detect distinguishing information will impact the interference. The presence or absence of interference due to an object can therefore be readily recorded in the visible photon channel, i.e. using photons that have not themselves interacted with the object. Crucially, the image transfer process leaves the thermal background behind, allowing for detection sensitivities that are considerably improved over direct IR detection \cite{YueThermal}. This principle has since been demonstrated for a variety of applications, including microscopy \cite{MicroscopyRamelow,BiologicalMicroscopy}, hyperspectral imaging\cite{paterova2020hyperspectral}, spectroscopy \cite{SpectroscopyPaterova,VisFTIR,NIRgratingRamelow,AgGaS2Paterova}, optical coherence tomography \cite{OCTPaterova,OCTRamelow20}, and holography \cite{HolographyGrafe}. 
	
IUP certainly offers a promising alternative to direct infrared imaging, but it is important to address the potential barriers to practical implementation, and a literature is emerging that tackles issues of size, stability, and speed \cite{PaterovaSemiconductor,VideoRateCompactGrafe,HolographyGrafe}, with compact near- and mid-IR technologies already seeing use in environmental and agricultural studies \cite{NIRHandheld,MIRHandheld}.

Here, we demonstrate two generations of compact, fully self-contained, wavelength-tunable, and low cost devices for IUP, both of which enable SWIR imaging using only a basic silicon CMOS camera. We also discuss a rapid analysis approach which uses a pixel-wise Fourier transform to extract both transmission and phase information from as few as three image frames. These developments have allowed us to make dramatic reductions to the size, weight, cost, and power (SWaP-C) of IUP.

\section{Methods}

\begin{figure}[ht!]
\centering\includegraphics[width=0.95\linewidth]{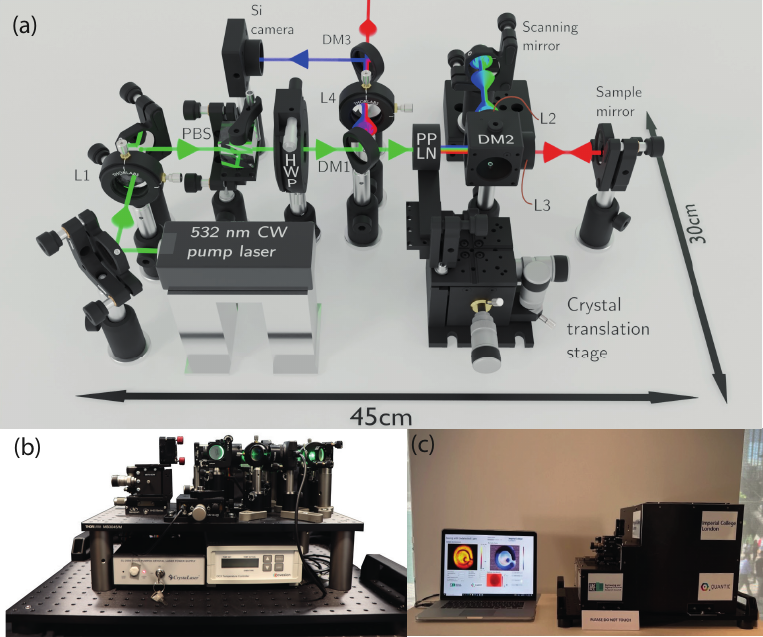}
		\caption{(a) Schematic of a compact setup for imaging with undetected photons. Green indicates the pump beam. Blue and red represent signal (visible) and idler (SWIR) beams, respectively. (b) Implementation of the compact device. The top breadboard (45\,$\times$\,30\,cm) contains the imaging system, with an additional 60\,$\times$\,45 cm breadboard to house system controls. (c) Demonstration of imaging with undetected photons outside of a laboratory, featuring system enclosure and real-time analysis output to a graphical user interface. }
		\label{fig:compact_v1}
	\end{figure}

Figure~\ref{fig:compact_v1}(a) shows the experimental setup of the first generation of our compact device. A 532\,nm diode-pumped solid-state continuous-wave laser (CrystaLaser CL532-050) pumps a periodically-poled lithium niobate (PPLN) crystal with 35\,mW of input power to produce signal (visible) and idler (IR) photons by SPDC. The dichroic mirror DM2 splits the idler photons from the signal and pump photons. The IR idler photons are sent to the sample, mounted on the sample mirror, while visible and pump photons propagate towards the scanning mirror, which is scanned to generate the interference fringes. All three wavelengths are then reflected back through the crystal, and as the pump makes its second pass, there is again a probability of generating a signal-idler photon pair. 

As previously discussed, if the optical modes from the second pass are perfectly overlapped with those from the first then a sinusoidal modulation appears in the signal photon flux detected at the camera as the scanning mirror is moved. At these powers, the actual probability of generating a photon pair at each pass is low, allowing us to neglect the possibility of any stimulated down-conversion in the second process by photons originating from the first.
	
An object placed on the sample mirror can introduce a loss and/or phase change to the idler from the first pair, introducing distinguishability and a proportional change in the amplitude and/or phase of the visible interference fringes. The amplitude variations can be imaged directly, but the phase changes are only detectable by moving the scanning mirror.

Both signal and idler photons are separated from the pump by dichroic mirror DM1 and sent to another dichroic mirror DM3, which sends only the visible photons to the camera. The idler photons go completely undetected. In fact, the silicon camera we use would not see them even if they were not removed by DM3.

To form the imaging system, each arm of the interferometer contains an $f\,=\,$50\,mm focal length lens (L2 and L3) such that both sample and scanning mirrors are in the image plane of the PPLN crystal, and the interferometer output is subsequently imaged onto the camera with a $f\,=\,$75\,mm focal length lens. A number of camera frames are recorded at different scanning mirror positions,  and the pixel-wise intensity variations are Fourier transformed \cite{THzFFT}, allowing the transmission and phase information of the object's IR response to be extracted.
	
The PPLN crystal can be translated perpendicular to the pump beam to access regions with different poling periods, in a way that tunes the signal and idler wavelengths. Also, it can be temperature tuned to further extend the overall wavelength coverage, allowing us to generate signal photons from 706-839 nm and idler photons from 1455-2159 nm.
	
The whole system sits within a 60 cm\,$\times$\,45 cm footprint, and is 40 cm high (including all the laser and the temperature controlling and scanning electronics, Figure \ref{fig:compact_v1}(b)). An additional enclosure can be added to reduce background light and allow safe operation outside of a lab environment, as demonstrated in Fig. \ref{fig:compact_v1}(c). In this case, the idler beam passes through an  AR-coated silicon window (which is opaque to both pump and signal)  to reach the sample which sits outside of the enclosure.  Typically, no realignment was required after local transportation and, once aligned, the system can be operated by an untrained user assuming familiarity with the image acquisition software. The whole system is assembled from standard off the shelf components for $\sim$£7000 (excluding the laser).

\section{Results}

\onecolumngrid

\begin{figure}[ht]
		\centering
		\includegraphics[width=0.95\linewidth]{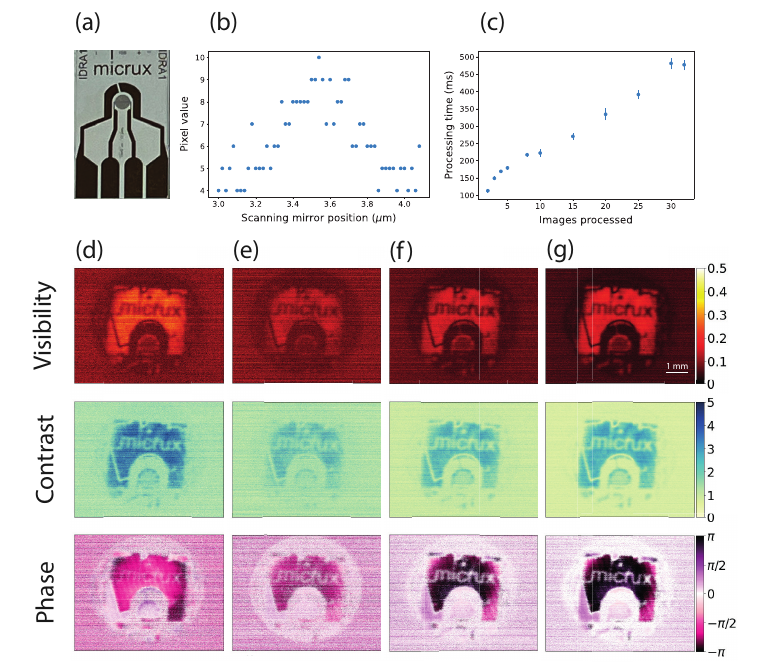}
		\caption{(a) Visible image of an epoxy/gold interdigitated ring array microelectrode used as a test sample. (b) Typical interference fringe profile seen at one pixel of the camera as the scanning mirror is moved. (c) Time taken to calculate contrast, visibility, and phase from varying numbers of input images\textemdash times are averaged over 100 runs with error bars given by the  standard deviation of these runs.  Visibility, contrast, and phase profiles of the microelectrode shown in (a),  generated from a pixel-wise Fourier transform of (d) 3, (e) 4, (f) 8, and (g) 15 images acquired while scanning the interferometer. Images were acquired with  200\,ms exposure.}
		\label{fig:images}
	\end{figure}
\twocolumngrid
Figure~\ref{fig:images} shows analysis of a thin-film gold interdigitated ring array microelectrode (Figure~\ref{fig:images}(a), Micrux IDRA1) using 3 (d), 4 (e), 8 (f), and 15 (g) acquired images. In each case, the images are taken with a 200\,ms exposure time at equally spaced piezo voltages over one oscillation of the interference pattern. The period of this oscillation is determined solely by the idler wavelength, as both signal and pump fields are scanned together, as shown in Figure \ref{fig:images}(b). The scanning mirror moves by close to half the idler wavelength, as the path is travelled twice. The detected wavelength is 808\,nm, which is filtered using a 10\,nm wide 810\,nm bandpass filter, while the probe wavelength is 1559\,nm. The images have a pixel size of 5.2\,$\times$\,5.2\,$\mu$m and are 1024\,$\times$\,1280 pixels, i.e. somewhat greater than those available with current IR cameras.

From the analysed images in Fig.~\ref{fig:images}(d), it is clear that both the phase and amplitude features of the sample can be identified reliably, even when working right at the Nyquist limit when as few as 3 recorded images are used for the analysis. This approach drastically reduces both acquisition and processing times. We define visibility as 
\begin{equation}
	\mathcal{V} = \frac{N_{max} - N_{min}}{N_{max} + N_{min}} = 2\frac{F_{1}}{F_{0}}
	\label{Eq:Vis}
\end{equation}
where $N_{max}$ ($N_{min}$) are the maximum (minimum) pixel values recorded on the camera during a phase scan, $F_{1}$ is the amplitude of the Fourier component which corresponds to the frequency of the interference oscillation, and $F_{0}$ is the amplitude of the DC Fourier component. Contrast is defined as 
\begin{equation}
	\mathcal{C} = N_{max} - N_{min} = F_{1} \,.
	\label{Eq:Contrast}
\end{equation}
Contrast is dependent on the overall brightness of a given system but can be a useful metric if there is a high detector noise floor as this will be subtracted, unlike the case with visibility.
Phase is defined as 
\begin{equation}
	\phi = \arctan\frac{\text{Im} (F_{1})}{\text{Re} (F_{1})} \, .
	\label{Eq:Phase}
\end{equation}
	
\noindent Although as few as 3 images are sufficient for a qualitative analysis of the relative phase and transmission across a sample, unsurprisingly, more images improve accuracy if further parameter extraction is desired. 
	
Features appear brighter in Figure~\ref{fig:images}(d) compared to Figure~\ref{fig:images}(e) due to where the frequency of the interference oscillation occurs on the Fourier transform sampling frequencies. Leakage into more than one Fourier component will be seen as a loss of signal and thus visibility will be reduced. This could be avoided by altering the Fourier transform length (via zero-padding) to better match the interference frequency to one of the sampled Fourier frequencies.
	
The time taken to perform the Fourier transform and calculate the above parameters is plotted against the number of input images in Figure~\ref{fig:images}(c). The Fourier transform is implemented using a Python wrapper of the Fastest Fourier Transform in the West (FFTW)\cite{FFTW,pyfftw} on a typical laboratory machine (Intel Xeon W-2102 processor, 4 cores). Processing times do not include saving or displaying the data. There is no delay between acquisitions required for image postprocessing  \cite{HolographyGrafe} and we require both fewer images and shorter exposure times than seen in Ref. \onlinecite{PaterovaSemiconductor}.

\begin{figure}[ht]
    \centering
    \includegraphics[width=0.95\linewidth]{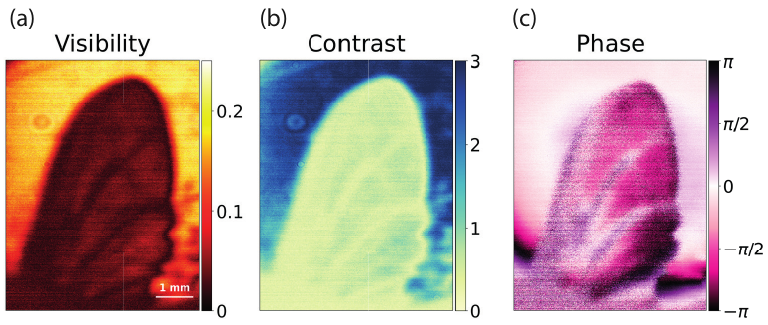}
    \caption{(a) Visibility, (b) contrast, and (c) phase images of a fly wing, probed at 1559\,nm and detected at 808\,nm.}
    \label{fig:FlyWing} 
    \end{figure}

Another example of SWIR imaging is shown in Figure \ref{fig:FlyWing}, this time with an organic sample of a fly wing. This provides an object with continuously varying IR transmission across the image, rather than the example of the electrode which only has regions of either total transmission or no transmission. Both samples shown thus far have been measured in transmission. Regions of high visibility are seen where SWIR light passes through the sample to be reflected by the sample mirror. This requires that the path from the crystal to the sample mirror and the path from the crystal to the scanning mirror must be equal to within the SPDC coherence length ($\approx$\,0.1\,mm). Any optical path length introduced by transmission through the sample must also be considered.
    
\begin{figure}[ht]
		\centering
		\includegraphics[width=0.95\linewidth]{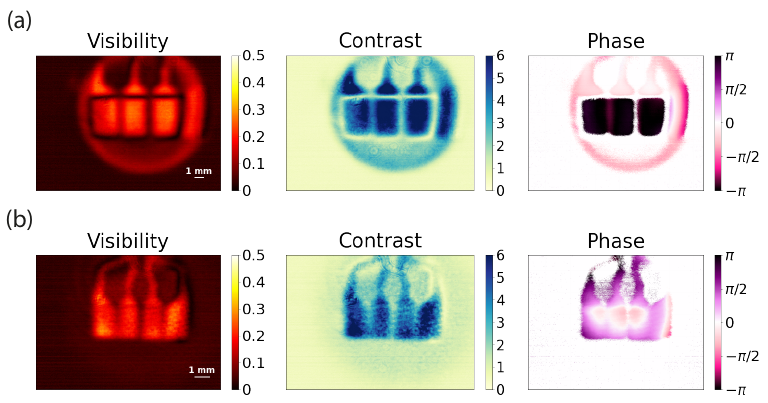}
		\caption{Visibility, contrast, and phase of gold electrodes at different probe wavelengths, reached by translating crystal and adjusting temperature. (a) Probe wavelength 1558\,nm and detected wavelength 808\,nm. (b) Probe wavelength 1818\,nm and detected wavelength 752\,nm.}
		\label{fig:longwl}
	\end{figure}

Figure \ref{fig:longwl} shows the wavelength tunability of the system while imaging the gold contacts of the electrode, shown in the bottom half of Figure \ref{fig:images}(a). In Fig.~\ref{fig:longwl}(a), the crystal is kept at 125$^{\circ}$C and the pump beam enters a region with a poling period of 7.40\,$\mu$m. These conditions result in a probe wavelength of 1558\,nm and a detected wavelength of 808\,nm (filtered with a 10\,nm wide 810\,nm bandpass). The crystal is then translated perpendicular to the pump beam to access a poling period of 7.71\,$\mu$m and heated to 200$^{\circ}$C. This extends the probe wavelength to 1818\,nm, beyond the sensitivity of a typical InGaAs camera, with the visible detection at 752\,nm (filtered with a 10\,nm wide 750\,nm bandpass). The interferometer remains aligned throughout these types of wavelength sweep.

Both the spatial resolution ($\Delta x = f_{u}\lambda_{u} / \sqrt{2}\pi w_p$) and magnification ($M = f_{c}\lambda_{d} / f_{c}\lambda_{d}$) are reduced as the idler probe wavelength increases. Here,  $f_u$ is the focal length of the lens in the undetected path, $f_c$ is the focal length of the lens in front of the camera, $\lambda_{d}$ is the detected wavelength, $\lambda_{u}$ is the undetected probe wavelength, and $w_p$ is the beam waist of the pump \cite{ResolutionMomentum}. The focal lengths of the lenses are also likely to vary from their nominal values as the wavelengths change. This change in magnification leads to all 4 contacts being visible in Figure \ref{fig:longwl}(b), while only 3 can be seen in Figure \ref{fig:longwl}(a), although chromatic aberrations in the lenses may also be limiting the resolution. Here the sample is being imaged in reflection, with high visibility in regions where the idler probe is reflected by the gold features, in which case it is the path from the crystal to the front face of the sample (rather than the sample mirror) that must be matched to the path from the crystal to the scanning mirror.
	
\section{Further Development: `EntangleCam'}

\begin{figure}[ht]
		\centering,
		\includegraphics[width=0.95\linewidth]{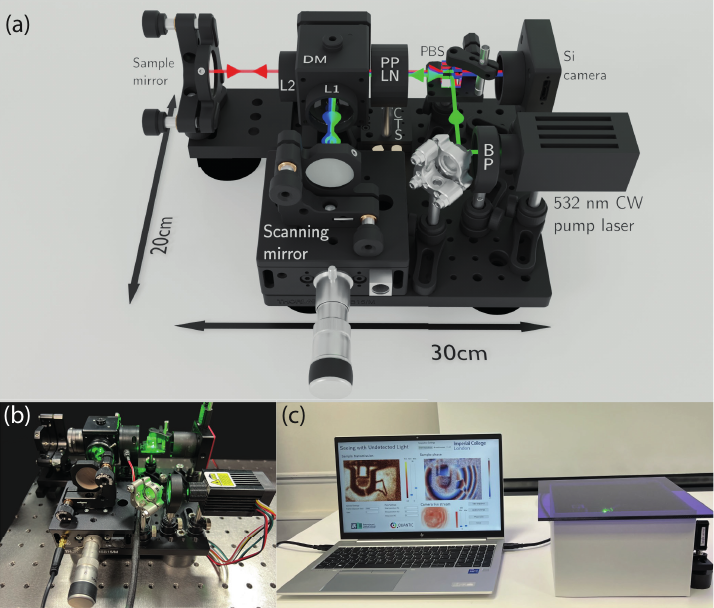}
		\caption{(a) Schematic of second generation of compact imaging with undetected photons, with further reduced SWaP-C. Green indicates the pump beam. Blue and red represent signal (visible) and idler (SWIR) beams, respectively. (b) Implementation of the second generation compact device. (c) Demonstration of the device outside of a laboratory, featuring system enclosure and real-time analysis output to a graphical user interface.}
		\label{fig:compact_v2}
	\end{figure}

Figure \ref{fig:compact_v2} shows the next design iteration of the system, the so-called `EntangleCam', using smaller optomechanics and simplifying the layout by removing and combining some of the components. The PBS that was used in the previous system to filter the pump polarisation has been replaced by a laser line PBS with an AR coating that doubles as a dichroic mirror to separate the signal from pump. The lens in front of the camera has been removed so that the detector array samples the far-field of the crystal directly, and  the original laser has been replaced by a simple 532\,nm laser diode (OdicForce green laser module) providing less than 30\,mW of light to the crystal. The pump shaping preparation is handled entirely by a single lens after the diode, with a bandpass filter (BP) to eliminate any unwanted IR light.

The breadboard footprint has shrunk to 30\,$\times$\,20\,cm$^2$ and is only 15\,cm high.The control electronics run from a single mains socket and add only 25\,$\times$\,22.5\,$\times$\,7.5\,cm$^2$ volume, while the total component cost is reduced to $\sim$£6,000, in this case including the laser.

\begin{figure}[ht]
		\centering
		\includegraphics[width=0.95\linewidth]{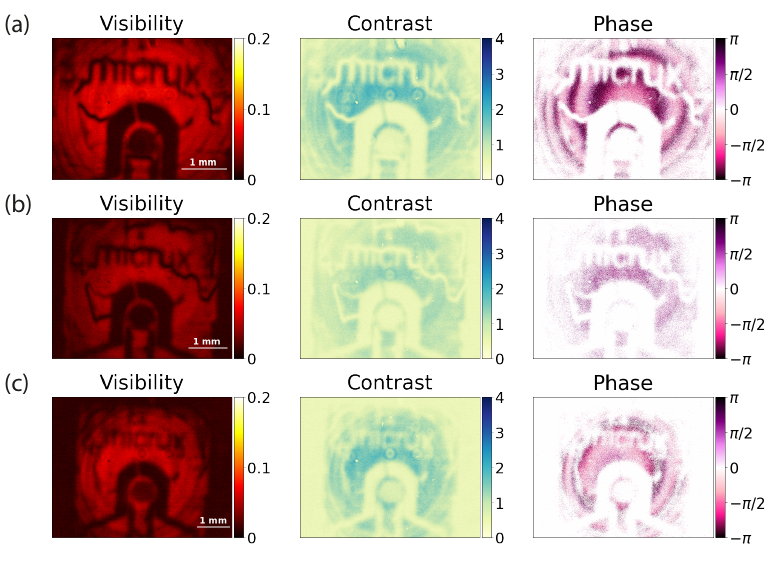}
		\caption{Visibility, contrast, and phase of gold microelectrode at different probe wavelengths using the system in Figure \ref{fig:compact_v2}. Different wavelengths are reached by translating crystal whilst leaving its temperature at room temperature. (a) Probe wavelength 1450\,nm and detected wavelength 840\,nm. (b) Probe wavelength 1620\,nm and detected wavelength 792\,nm. (c) Probe wavelength 1783\,nm and detected wavelength 758\,nm.}
		\label{fig:v2_longwl}
	\end{figure}

 The system still retains the same wavelength tunability via crystal heating and translation with crystal translation stage (CTS), but even at a fixed room temperature, significant wavelength tunability is available. This is seen in Figure \ref{fig:v2_longwl}, imaging another gold electrode sample (identical to that seen in Figure \ref{fig:images}), using only a room temperature crystal (24.5$^{\circ}$C) without active temperature control. The bandpass filter on the camera was replaced with a long-pass filter to reject any residual pump light while also removing the need to change detection filters when moving between different poling periods. Images used in this analysis are taken with a 200\,ms exposure time. It can be seen that despite the reduction in size, the system retains its imaging capabilities and broad tunability. Depending on the probe wavelength desired for a particular application, this shows that the device could be designed to operate without any need for temperature control, further reducing SWaP-C.

 \section{Discussion}

In conclusion, we have demonstrated compact IUP systems that can perform infrared imaging with visible detection and are robust and portable enough to be used outside of a laboratory environment. We have also demonstrated rapid analysis that allows a real-time quantitative measure of transmission and phase shift of a sample using as few as 3 recorded images. These systems represent a significant step forwards in the affordability and practicality of IUP as an alternative to direct IR imaging.
	
Future operation speeds could be enhanced by reducing the acquisition time required at each scanning mirror position, simply by using a higher power laser and/or a more sensitive camera. Both are readily available without major cost implications. The spatial resolution of the system can be further enhanced by increasing the momentum correlations between the signal and idler photons; in the current configuration this would be achieved by increasing the width of the pump beam \cite{ResolutionMomentum}. 
    
Furthermore, there are a number of ways in which the operating wavelengths can be extended towards the mid-IR, including different pump wavelengths \cite{NathanBalancing}, different poling periods \cite{XtalSuperlattice,Aperiodic}, and different nonlinear materials \cite{AgGaS2,AgGaS2Paterova,TuneableFiber}. By moving towards the mid-IR, we anticipate our system will be a valuable tool for chemically and medically relevant applications \cite{MolecularMixtures,OralCancer,IRcancerreview,Digistain}. 

\begin{acknowledgments}
We acknowledge funding from the UK National Quantum Hub for Imaging (QUANTIC, No. EP/T00097X/1), an EPSRC DTP, and the Royal Society (No. UF160475). The authors declare no conflicts of interest. 
\end{acknowledgments}

\bibliography{Compact}

\end{document}